\documentclass[floatfix,%
 reprint,%
 amssymb, amsmath,%
 aip,jcp,
frontmatterverbose,
]{revtex4-2}

\usepackage{amsmath,amssymb,amsfonts,mathtools}
\usepackage{multirow}
\usepackage{float}
\usepackage{csquotes}
\usepackage{dsfont}
\usepackage[mathscr]{euscript}
\usepackage{natbib}

\usepackage{enumitem}
\setlist{nosep} 

\usepackage{graphicx}
\usepackage{bm}%
\usepackage{longtable}
\usepackage{supertabular}
\usepackage{rotating}
\usepackage[normalem]{ulem}
\usepackage{ifthen}
\newboolean{includeMQDT}
\setboolean{includeMQDT}{false}  

\usepackage[dvipsnames]{xcolor}

\def\measurehat#1{%
   \setbox0=\vbox{$\hat{#1}\hfil\break$\null\par
      \setbox0=\lastbox\unskip\unpenalty\global\setbox1=\lastbox}%
   \setbox0=\hbox{\unhbox1 \unskip\unpenalty\unskip \global\setbox2=\lastbox}%
   \setbox0=\vbox{\unvbox2 \setbox0=\lastbox}%
}
\newcommand{\doublehat}[1]{%
   \measurehat{#1}\dimen0=\wd0 \measurehat{\kern0pt{#1}}%
   \raise.35ex\rlap{\kern\dimexpr\dimen0-\wd0$\hat{\phantom{#1}}$}{\hat{#1}}%
}

\usepackage{dcolumn,ulem}  
\usepackage{xspace}

\begin{document}

\title{\large Spin relaxation dynamics with a continuous spin environment: the dissipaton equation of motion approach \vspace{0.25cm}}

\author{Wenxiang Ying}
\email{wying3@ur.rochester.edu}
\affiliation{Department of Chemistry, University of Rochester, 120 Trustee Road, Rochester, New York 14627, USA \vspace{0.25cm}}
\author{Yu Su}
\affiliation{Department of Chemical Physics, University of Science and Technology of China, Hefei, Anhui 230026, China}
\author{Zi-Hao Chen}
\affiliation{Department of Chemical Physics, University of Science and Technology of China, Hefei, Anhui 230026, China}
\author{Yao Wang}
\email{wy2010@ustc.edu.cn}
\affiliation{Department of Chemical Physics, University of Science and Technology of China, Hefei, Anhui 230026, China}
\author{Pengfei Huo}

\email{pengfei.huo@rochester.edu}
\affiliation{Department of Chemistry, University of Rochester, 120 Trustee Road, Rochester, New York 14627, USA \vspace{0.25cm}
\affiliation{The Institute of Optics, Hajim School of Engineering, University of Rochester, Rochester, New York 14627, USA}
}

\date{\today}

\begin{abstract}

We present the quantum dynamics of a spin coupling to a bath of independent spins via the dissipaton equation of motion (DEOM) approach. The bath, characterized by a continuous spectral density function, is composed of spins that are independent level systems described by the $\mathfrak{su}(2)$ Lie algebra. This represents an extreme class of anharmonic environment. 
Based on the conclusion drawn by Suarez and Silbey [J. Chem. Phys. 95, 9115 (1991)] and Makri [J. Chem. Phys. 111, 6164 (1999)] that the spin bath can be mapped to a Gaussian environment under its linear response limit, we derive the fluctuation-dissipation theorem (FDT) of the spin bath from a microscopic perspective, and generalize the discussion to the case of arbitrary bath spin quantum number $S$. Next, the time-domain Prony fitting decomposition scheme is applied to the bare-bath time correlation function (TCF) given by FDT to generate the exponential decay basis (or pseudo modes) for DEOM construction. The accuracy and efficiency of this strategy has been justified by a variety of numerical results. We envision this work provides new insights to extend the hierarchical equations of motion (HEOM) and DEOM approach to certain types of anharmonic enviroments with arbitrary TCF or spectral density. 

\end{abstract}

\maketitle

\section{Introduction} \label{sec:level1}

The dynamics of a two-level-system (TLS) coupling to a dissipative environment has been extensively investigated over the past decades. The most widely and systematically studied one is the spin-boson system,\cite{Ohmic_0, Weiss, Nitzan, Hanggi_1990} which has various applications in physics and chemical dynamics in condensed phase\cite{Ohmic_0, Weiss, Nitzan, Hanggi_1990, Kofman_2004, PCET_2010}. The environmental part of the spin-boson system is a set of non-interacting harmonic oscillators. This is reasonable as Caldeira and Leggett \cite{Caldeira_Leggett_1983} had justified the universality of bosonic heat baths consisting of an infinite number of harmonic oscillators which are linearly coupled to the system in the real physical world. 

Apart from the bosonic environment, another typical environment of interest is a bath consisting of a set of spins, \cite{Caldeira_1993, Yan_2014} which can be regarded as an extreme example of anharmonic environment. The spin-spin-bath (SSB) model denotes a TLS coupled with a dissipative spin bath. It has recently drawn tremendous attention due to the gradually accumulating phenomena in physical setups. Here are some evidences. It is predicted and justified that at very low temperature, the dynamics of magnetic nanomolecules, such as $\mathrm{Fe}_8$, $\mathrm{Mn}_{12}$, is strongly influenced by nuclear spins; \cite{LowT_1998, LowT_exp1, LowT_exp2, LowT_exp3, Nuclear_2003} for solid-state quantum computing devices whose qubits are typically electron spins, such as GaAs quantum dot \cite{Steane_1998, Liu_2005, QPT_1} and diamonds with nitrogen-vacancy (NV) centres \cite{Liu_2009, Liu_2012}, it is inevitable for the qubits to be coupled to environmental spins. The spin environment is also concerned in stylized quantum measurement setups,  \cite{zurek_1991, Zurek_1997} the studies of quantum phase transition, \cite{QPT_1, QPT_2, QPT_3} and more recently, the radical pair spin relaxation as well as its applications in quantum information processing, \cite{Manolopoulos_2019, Manolopoulos_2020, JACS_2021, Science_2020} etc. Anharmonic but very simple, the SSB model provides new insights for quantum dynamics in condensed phase. 

The theoretical treatment to the anharmonic spin bath has first been discussed by Suarez and Silbey \cite{Suarez_Silbey_1991}, Makri \cite{Makri_JCP1999}, and later by Yan \cite{YYA_2016}. They proved that under the thermodynamics limit (also known as the linear response limit), the generally anharmonic spin bath approximately satisfies the Gaussian statistics. As a result, the spin bath can be effectively characterized by the familiar boson bath with a modified spectral density function, which is much easier to be treated by a wide range of quantum dynamics methods. On this basis, higher order nonlinear effects can be further studied by considering a finite number of bath spins. Previous work on the quantum dynamics of SSB types of models have been carried out extensively by iterative path integral approach based on influence functional, \cite{Makri_JCP1999, Makri_JPCB1999, Nancy_PCCP_2018} the multilayer multiconfiguration time-dependent Hartree (ML-MCTDH) \cite{Wang_Shao_2012} and the closely related surrogate Hamiltonian \cite{SSH_2004, SSH_2019} approach, polaron-transformed master equation, \cite{NingYang_2013} Nakajima-Zwanzig type of generalized master equation,\cite{Sarma_2012} and the generalized hierarchical equations of motion (gHEOM) that based on the stochastic Liouville equation with perturbative expansion scheme, \cite{Cao_2018_1, Cao_2018_2} etc. 

A variety of approximation-based methods are also developed and applied for the spin bath problems, such as the time-dependent perturbation approach,\cite{Hang_2009} phase space quasi-classical methods, \cite{Pechukas_1999} etc. A number of cluster expansion (CE) approaches \cite{Sarma_2007}, including the cluster correlation expansion \cite{Liu_2006, Liu_2008, YANG_2020}, linked cluster expansion \cite{Sham_2007}, and the associated dynamical mean field theory \cite{Timo_2021} have been developed to study the many-body bath time evolution. Nakamura and Tanimura \cite{Tanimura_2021} studied the dynamics of a TLS that interacts with a subenvironment consisting of a one-dimensional \textit{XXZ} spin chain using the hierarchical Schr{\"o}dinger equations of motion (HSEOM), despite that the noise generated from the spin lattice is non-Gaussian and non-local. \cite{Weiss, Kleinert} The series of work done by Fay, Lindoy, and Manolopoulos \cite{Lachlan_2018, Lachlan_JCP_2018, Manolopoulos_2019, Manolopoulos_2020, Lachlan_2020, Lachlan_2021} had provided a systematic theoretical framework for electron spin relaxation in radical pairs that exposed to an environment of nuclear spins, which includes full quantum mechanical treatment of all spin degrees of freedom using tensor network propagation strategy, and master equation approaches based on the Schulten-Wolynes semiclassical treatment to the nuclear spins. 

Despite the fruitful progresses, these approaches all have their specific limitations. For example, the direct path integral-based method are numerically efficient only for short memory length; \cite{Makri_1994, Makri_1995, Reichman_2012} the numerically exact ML-MCTDH approach relies on a discretization strategy for the continuous bath spectral density, is computational costly to reach numerical convergence for models with high bath cutoff frequency, so that Born-Oppenheimer (B-O) type of approximation needs to be made; \cite{Ohmic_0, Wang_Shao_2012} MCTDH is even more computational costly under the finite-temperature case due to its dependence on the Monte Carlo sampling strategy. \cite{Meyer_MCTDH} Besides, almost all of the work mentioned above only take care of the spin-1/2 case. In realistic situations, however, the nuclear spin quantum numbers are usually much larger than $1/2$.

Regarding to the limitations of current studies on the spin bath that mentioned above, a more general, numerically efficient and accurate theoretical framework needs to be developed. The dissipaton equation of motion (DEOM) is a statistical quasi-particle theory for quantum dissipative dynamics. Featured by the very powerful dissipaton algebras, it not only just recovers HEOM formalism, \cite{Tanimura_1990, JPSoc_1990, Rxxu_2005} but also identifies the physical meanings of the dynamical variables. \cite{Yan_2014, Yan_2016} Equipped with the time-domain Prony fitting decomposition ($t$-PFD) \cite{PFD_2022} scheme, by which the environmental TCF can be accurately decomposed into exponential sums, the efficiency and applicability of HEOM/DEOM is significantly improved. It is notable that $t$-PFD scheme is, in principle, applicable to arbitrary bath TCF or spectral density. Under the linear response limit, the theory of effective spectral density function provides a perfect platform for us to play with $t$-PFD. 

In this work, by taking advantage of the linear response limit, we generalize the theory of effective spectral density function for spin bath with arbitrary bath spin quantum number $S$. Then we apply DEOM to study the spin relaxation dynamics of a TLS coupling to the effective \textit{bosonic} environment. The bath TCF decomposition is achieved by $t$-PFD. This paper is organized as follows. In Section \ref{section_II}, we briefly review the spin bath model, its linear response properties, bosonic DEOM, and $t$-PFD scheme; in Section \ref{section_III}, we present various numerical testing results for SSB models, including zero- and finite-temperature, weak and strong coupling, unbiased and biased cases with comparisons against ML-MCTDH; in Section \ref{section_IV}, we briefly summarized the major advantage of our method.

\section{Model and methodology} \label{section_II}

The total TLS-plus-bath composite Hamiltonian of an open quantum system reads as
\begin{align}
    \hat{H} = \hat{H}_{\mathrm{S}} + \hat{h}_{\textsc{B}} + \hat{H}_{\mathrm{SB}},
\end{align}
where $\hat{H}_{\mathrm{S}} \equiv \epsilon \hat{\sigma}_z + \Delta \hat{\sigma}_x$ is the system Hamiltonian of a TLS with energy bias $\epsilon$ and off-diagonal coupling $\Delta$, $\hat{\bm{\sigma}} \equiv (\hat{\sigma}_x, \hat{\sigma}_y, \hat{\sigma}_z)$ denotes to the Pauli matrices; $\hat{h}_{\textsc{B}}$ is the bath Hamiltonian which usually consists of a macroscopic number of noninteracting particles that can be bosons, fermions and/or spins. $\hat{H}_{\mathrm{SB}}$ carries the system-bath interaction. The influence of the bath entails quantum statistical mechanics description, in which the thermodynamics limit is naturally assumed. We set $\hbar \equiv 1$ throughout this paper for the sake of convenience. 

\subsection{The spin bath models}

The spin bath and its interaction with the system can be described by the Hamiltonian below, \cite{Makri_JCP1999, Wang_Shao_2012, Cao_2018_2}
\begin{subequations}
    \begin{align}
        \hat{h}_{\textsc{B}} &=  \sum_{j = 1}^N \omega_j \hat{s}^j_z, \label{bathHam}\\
        \hat{H}_{\mathrm{SB}} &= \sum_{a} \hat{Q}_a \otimes \hat{F}_a.
    \end{align}
\end{subequations}
The bath Hamiltonian $\hat{h}_{\textsc{B}}$ describes $N$ independent spins which are distinguishable. It is diagonal in the $\{ (\hat{\bm s}^j)^2, \hat{s}^j_z \ |\ j = 1, \cdots, N\}$ eigen representation, where $\hat{\bm{s}}^j$ are the spin operators associated with the $j$-th bath mode, characterized by energy difference $\omega_j$. They form the $\otimes_{j = 1}^N \mathfrak{su}(2)$ Lie algebra, $[\hat{s}^i_{\alpha}, \hat{s}^j_{\beta}] = i \delta_{ij} \epsilon_{\alpha \beta \gamma} \hat{s}^i_{\gamma}$, where $\alpha, \beta, \gamma$ denotes to the Cartesian components of the spin matrices, $\epsilon_{\alpha \beta \gamma}$ is the 3-D Levi-Civita tensor, $\delta_{ij}$ is the Kronecker symbol. $\{ \hat{Q}_a \}$ and $\{ \hat{F}_a \}$ denotes to the system and bath dissipation modes, respectively.

There are extensive types of spin-spin interaction, such as the Ising type, \cite{Ising_1925} the Heisenberg type, \cite{Dirac_1929} etc., whose general expression takes the form $\hat{\sigma}_{\alpha} \hat{s}^j_{\beta}$. Among the various choices, the most common types can be summarized as \cite{Cao_2018_2}
\begin{align} \label{HSB_choices}
    \hat{H}_{\mathrm{SB}} = \begin{cases}
        \frac{1}{\sqrt{S}} \sum_{j = 1}^N c_j \hat{s}^j_x \hat{\sigma}_z, \vspace{0.15cm}\\ 
        \frac{1}{\sqrt{S}} \sum_{j = 1}^N c_j \hat{s}^j_z \hat{\sigma}_z, \vspace{0.15cm}\\ 
        \frac{1}{\sqrt{S}} \sum_{j = 1}^N c_j (\hat{s}^j_x \hat{\sigma}_x + \hat{s}^j_y \hat{\sigma}_y), \vspace{0.15cm} \\
        \frac{1}{\sqrt{S}} \sum_{j = 1}^N c_j(\hat{s}^j_x \hat{\sigma}_x + \hat{s}^j_y \hat{\sigma}_y + \hat{s}^j_z \hat{\sigma}_z),
    \end{cases}
\end{align}
where $c_j$ are the coupling coefficients between the system dissipation operators and the $j$-th bath mode, $S$ is the spin quantum number of the bath spins. In this paper, we shall mainly focus on the first interaction form in Eq.(\ref{HSB_choices}). According to Caldeira and Leggett, \cite{Caldeira_Leggett_1983}  the bath as well as its coupling to the system might be described by the spectral density function, defined as 
\begin{align} \label{spectral_density_def_original}
    J_{ab}(\omega) &\equiv \frac{\pi}{2} \sum_{j = 1}^N c^*_{aj} c_{bj} \delta(\omega - \omega_j),
\end{align}
where the coupling coefficients are assumed to obey the general scaling rule: $c_{aj} \sim 1 / \sqrt{N}$. 

It is worth noting that the spin bath is not generally a Gaussian environment. Suarez and Silbey \cite{Suarez_Silbey_1991}, Makri \cite{Makri_JCP1999} have remarkably shown that under the limit of $N \rightarrow + \infty$, the SSB model with the first interaction form in Eq.(\ref{HSB_choices}) and bath spin quantum number $S = 1/2$, can be rigorously mapped onto the familiar spin-boson model with an effective spectral density,
\begin{align}
    [J_{ab}]_{\mathrm{eff}} (\omega; \beta) = J_{ab} (\omega) \tanh (\beta \omega / 2), \label{Jeff}
\end{align}
with the subscript $a = b$ for single mode case. More specifically, only the second order term remains in the cumulant expansion of the influence functional as $N \rightarrow + \infty$. The higher order cumulants $\mathcal{O}(1/\sqrt{N})$ will disappear under the scaling limit. As a result, the noise spectrum of the bath dissipation operators are Gaussians, \cite{Suarez_Silbey_1991} leading to the linear response limit. \footnote{There is a straightforward example with respect to the pure dephasing case, in which $\Delta = 0$ such that the dephasing dynamics has analytical solution, as discussed by Rao and Kurizki \cite{Pure_dephasing}, Hsieh and Cao \cite{Cao_2018_2}}

\subsection{Linear response of the spin bath, and the generalized theory of effective spectral density function} \label{linear_response_section}

As is discussed above that the continuous spin bath approximately satisfies Gaussian statistics. The influence of Gaussian environments are completely characterized with the linear response functions of hybrid bath modes in the isolated bare-bath subspace. For bosonic environment, it can be defined via the commutator as
\begin{align} \label{fermionic_response_function}
    \chi_{ab} (t) \equiv i \langle [ \hat{F}_a(t), \hat{F}_b (0) ] \rangle_{\textsc{B}}.
\end{align}
Here $\hat{F}(t) = e^{i \hat{h}_{\textsc{B}} t} \hat{F}(0) e^{-i \hat{h}_{\textsc{B}} t}$ exerts the \textit{stochastic force}, and $\langle (\cdot) \rangle_{\textsc{B}} \equiv \mathrm{tr}_{\textsc{B}} [ (\cdot) \rho^{eq}_{\textsc{B}}(T) ]$ denotes ensemble average in the bath subspace, with $\rho^{eq}_{\textsc{B}}(T) \equiv e^{ - \beta \hat{h}_{\textsc{B}}} / \mathrm{tr}_{\textsc{B}} [e^{ - \beta \hat{h}_{\textsc{B}}}] $. For fermionic environments, the similar concept can also be defined via the anti-commutator, \cite{Weiss, Yan_2014} $G_{ab}(t) \equiv \langle \{ \hat{F}_a(t), \hat{F}_b (0) \} \rangle_{\textsc{B}}$, which is known as the single-particle Green's function. The \textit{causality} Fourier transform of $\chi_{ab}(t)$ and $G_{ab}(t)$ is defined as
\begin{subequations}
    \begin{align}
    \chi_{ab} (\omega) \equiv \int_0^{\infty}dt\ e^{i\omega t} i \langle [ \hat{F}_a(t), \hat{F}_b (0) ] \rangle_{\textsc{B}}, \\
    G_{ab} (\omega) \equiv \int_0^{\infty}dt\ e^{i\omega t} \langle \{ \hat{F}_a(t), \hat{F}_b (0) \} \rangle_{\textsc{B}}.
\end{align}
\end{subequations}
One can easily check their symmetry. 
The spectral density functions can be evaluated using time-reversal symmetry (TRS) as \cite{Weiss}
\begin{subequations}
    \begin{align} \label{spectral_density_def}
    J_{ab}(\omega) &\equiv \mathrm{Im}[\chi_{ab} (\omega)] = \frac{1}{2} \int_{- \infty}^{+\infty} dt\ e^{i \omega t} \langle [ \hat{F}_a(t), \hat{F}_b (0) ] \rangle_{\textsc{B}}, \\
    J'_{ab}(\omega) &\equiv \mathrm{Re}[G_{ab} (\omega)] = \frac{1}{2} \int_{- \infty}^{+\infty} dt\ e^{i \omega t} \langle \{ \hat{F}_a(t), \hat{F}_b (0) \} \rangle_{\textsc{B}}, \label{fermionic_Jdef}
\end{align}
\end{subequations}
corresponding to the bosonic and fermionic cases, respectively. Likewise, one can define the TCF as well as its spectrum functions as
\begin{subequations}
    \begin{align}
    C_{ab}(t) &\equiv \langle \hat{F}_a(t) \hat{F}_b (0) \rangle_{\textsc{B}}, \\
    C_{ab} (\omega) &\equiv \int_0^{\infty} dt\ e^{i\omega t} \langle \hat{F}_a(t) \hat{F}_b (0) \rangle_{\textsc{B}}. \label{corr_spectrum}
\end{align}
\end{subequations}

Without loss of generality, we consider the case of linear system-bath coupling with only one dissipation mode, 
\begin{align}
    \hat{F} \equiv \frac{1}{\sqrt{S}} \sum_j c_{j} \hat{s}^j_x,
\end{align}
for the spin bath with $S = 1/2$. Here we drop all the subscripts for simplicity. One will finally arrive at its FDT with respect to the auto-TCF of the bath dissipation operator,
\begin{align}
    C(t) = \frac{1}{\pi} \int_{- \infty}^{+\infty} d \omega \frac{e^{- i \omega t} J'(\omega) }{1 + e^{- \beta \omega}}. \label{fermionic_FDT}
\end{align}
See Appendix \ref{appendix_B} for detailed derivations. On the other hand, one can independently obtain that:
\begin{align} \label{bosonic_FDT}
    C(t) = \frac{1}{\pi} \int_{- \infty}^{+\infty} d\omega\ \frac{e^{- i \omega t} J_{\mathrm{eff}} (\omega; \beta) }{1 - e^{- \beta \omega}},
\end{align}
where $J_{\mathrm{eff}} (\omega; \beta) \equiv J(\omega) = J'(\omega) \tanh (\beta \omega / 2)$, recovering Eq.(\ref{Jeff}), the well-known result. So that very interestingly, the continuous spin environment is isomorphic to a boson environment with temperature-dependent effective spectral density function, whose zero-temperature limit gives rise to $J'(\omega) = J(\omega)$. For this reason, the zero-temperature spin-boson model is also widely known as \textit{taking the spin-bath limit}. \footnote{ Note that historically, Schotte\cite{Schotte_1970} first pointed out that at low temperature and long time limit ($t \gg \omega_c^{-1}$), the dynamics given by the Kondo Hamiltonian in a bosonic picture is the same as that of the corresponding fermionic operators. This pattern also emerges in the theoretical validation of the Caldeira-Leggett model \cite{Caldeira_Leggett_1983, Caldeira_1993} } 

Due to the generality of the linear response limit, the theory of effective spectral density is widely applicable, so that it is not restricted to $S = 1/2$ case, but with arbitrary $S$, as long as the system-bath coupling is linear. We also provide discussions upon arbitrary spin $S$ case in the Appendix \ref{appendix_B}, which can be viewed as a generalization of the effective spectral density function theory.

\subsection{Bosonic DEOM formalism}

Based on the previous discussion, the open quantum system problem with a continuous spin environment will be exactly mapped to the familiar spin-boson type of model with an effective spectral density function, which has the Hamiltonian description as below,
\begin{align}
    \hat{H} = \hat{H}_{\mathrm{S}} + \frac{1}{2} \sum_j \omega_j (\hat{x}^2_j + \hat{p}^2_j) + \sum_a \hat{Q}_{a} \otimes \sum_j c'_{aj} \hat{x}_{aj}, 
\end{align}
where $\hat{x}_j$, $\hat{p}_j$ are the conjugated coordinate-momentum pairs that satisfy the Heisenberg commutation relations, $\{ \hat{Q}_{a} \}$ are the original system dissipation mode, and $\{ c'_{aj} \}$ are the rescaled coupling coefficients due to the effective spectral density. The problem is now readily to be solved by HEOM/DEOM.

The DEOM theory is a statistical quasi-particle theory for quantum dissipative dynamics, describing the influence of bulk environments using only a few number of quasi-particles, the dissipatons. \cite{Yan_2014, Yan_2016} They arise strictly from the linear bath coupling component:
\begin{align} \label{dissipaton_decomp}
    \hat{F}_a = \sum_{k = 1}^{K} \hat{f}_{ak},
\end{align}
with single-damping parameters given by
\begin{subequations} \label{dissipaton_correlation}
    \begin{align}
    \langle \hat{f}_{ak} (t) \hat{f}_{bj} (0) \rangle_{\textsc{B}} = \delta_{kj} \eta_{abk} e^{- \gamma_{ak} t}, \label{dissipaton1} \\
    \langle \hat{f}_{bj} (0) \hat{f}_{ak} (t) \rangle_{\textsc{B}} = \delta_{kj} \eta_{ab\overline{k}}^* e^{- \gamma_{ak} t}. \label{dissipaton2}
\end{align}
\end{subequations}
The associated index $\overline{k}$ in Eq.(\ref{dissipaton2}) is defined as $\gamma_{a\overline{k}} = \gamma_{ak}^*$ to preserve TRS. Further denote 
\begin{subequations}
    \begin{align}
    \langle \hat{f}_{ak} \hat{f}_{bj} \rangle^>_{\textsc{B}} \equiv \langle \hat{f}_{ak} (0^+) \hat{f}_{bj} (0) \rangle_{\textsc{B}} = \delta_{kj} \eta_{abk}, \\
    \langle \hat{f}_{bj} \hat{f}_{ak} \rangle^<_{\textsc{B}} \equiv \langle \hat{f}_{bj} (0) \hat{f}_{ak} (0^+) \rangle_{\textsc{B}} = \delta_{kj} \eta_{ab\overline{k}}^*,
\end{align}
\end{subequations}
for later use in the dissipaton algebra. Note that they are different from $\langle \hat{f}_{ak} \hat{f}_{bj} \rangle_{\textsc{B}}$. Eq.(\ref{dissipaton_decomp}) and (\ref{dissipaton_correlation}) leads to
\begin{align}
    \langle \hat{F}_a(t) \hat{F}_b(0) \rangle_{\textsc{B}} = \sum_{k = 1}^{K} \eta_{abk} e^{- \gamma_{ak} t},
\end{align}
and its complex conjugation. 

The dynamical variables in DEOM are the dissipaton density operators (DDOs): 
\begin{align}\label{DDO}
    \rho^{(n)}_{\mathbf{n}} (t) \equiv \mathrm{Tr}_{\textsc{B}} \left[ \Big( \prod_{ak} \hat{f}^{n_{ak}}_{ak}  \Big)^\circ \rho_{\mathrm{T}} (t) \right],
\end{align}
where $\rho_{\mathrm{T}} (t)$ is the time-dependent total density matrix, the product of dissipatons inside $(\cdots)^\circ$ means \textit{irreducible}. And $(\mathrm{c \text{-} number})^\circ = 0$. Bosonic dissipatons are symmetric under permutation, $(\hat{f}_{ak} \hat{f}_{bj} )^\circ = (\hat{f}_{bj} \hat{f}_{ak})^\circ$. Each DDO in Eq.(\ref{DDO}) represents a specific configuration of $\mathbf{n} \equiv \{\cdots, n_{ak}, \cdots |\ a = 1, \cdots, M;\ k = 1, \cdots, K \}$, with $n = \sum_{ak} n_{ak}$ dissipatons in total (\textit{i.e.}, the number of tiers). We also denote that the associated DDO's index $\bm{\mathrm{n}}^{\pm}_{ak}$ differs from $\bm{\mathrm{n}}$ at the specified $n_{ak}$ by $\pm 1$, which means $n_{ak}$ is replaced by $n_{ak} \pm 1$. 

The DEOM formalism can be constructed according to the dissipaton algebra, which includes the \textit{generalized diffusion equation} and \textit{generalized Wick's theorem}. The generalized diffusion equation arises from the single-damping character in Eq.(\ref{dissipaton_correlation}), that
\begin{align} \label{GDE1}
    \mathrm{Tr}_{\textsc{B}} \left[ \left( \frac{\partial \hat{f}_{ak}}{\partial t} \right)_{\textsc{B}} \rho_{\mathrm{T}} (t) \right] = - \gamma_{ak} \mathrm{Tr}_{\textsc{B}} \left[ \hat{f}_{ak} \rho_{\mathrm{T}} (t) \right].
\end{align}
The generalized diffusion equation is applicable for the $\hat{h}_{\textsc{B}}$-action,
\begin{align} \label{GDE2}
    \rho^{(n)}_{\bm{\mathrm{n}}}(t; h^{\times}_{\textsc{B}}) &\equiv \mathrm{Tr}_{\textsc{B}} \left\{ \Big(\prod_{ak} \hat{f}^{n_{ak}}_{ak} \Big)^\circ [\hat{h}_{\textsc{B}}, \rho_{\mathrm{T}} (t)] \right\} \notag\\
    &= \mathrm{Tr}_{\textsc{B}} \left\{ \Big[ \Big(\prod_{ak} \hat{f}^{n_{ak}}_{ak} \Big)^\circ , \hat{h}_{\textsc{B}} \Big] \ \rho_{\mathrm{T}} (t) \right\} \notag\\
    &= - i \left( \sum_{ak} n_{ak} \gamma_{ak} \right) \rho^{(n)}_{\bm{\mathrm{n}}}(t),
\end{align}
where $h^{\times}_{\textsc{B}}\ \cdot \equiv [\hat{h}_{\textsc{B}},\ \cdot ]$, the second line of Eq.(\ref{GDE2}) arises from the equivalence between the Schr{\"o}dinger and Heisenberg prescription, and the last line goes with Heisenberg equations of motion, $(\partial \hat{f}_{ak} / \partial t)_{\textsc{B}} = - i[\hat{f}_{ak}, \hat{h}_{\textsc{B}}]$. Eq.(\ref{GDE2}) summarizes the contribution by the bath Hamiltonian to the DDOs dynamics.\cite{Yan_2014, Yan_2016}

The generalized Wick's theorem deals with the system-hybrid-bath interaction, reading as
\begin{subequations}
    \begin{align} \label{GWT}
    \mathrm{Tr}_{\textsc{B}} \left[ \Big(\prod_{ak} \hat{f}^{n_{ak}}_{ak} \Big)^\circ \hat{f}_{bj} \rho_{\mathrm{T}} (t) \right] & = \rho^{(n+1)}_{\bm{\mathrm{n}}^{+}_{bj}}(t) \\
    &\ \ \ + \sum_{ak} n_{ak} \langle \hat{f}_{ak} \hat{f}_{bj} \rangle^>_{\textsc{B}} \notag \rho^{(n-1)}_{\bm{\mathrm{n}}^{-}_{ak}}(t), \notag\\
    \mathrm{Tr}_{\textsc{B}} \left[ \Big(\prod_{ak} \hat{f}^{n_{ak}}_{ak} \Big)^\circ \rho_{\mathrm{T}} (t) \hat{f}_{bj} \right] & = \rho^{(n+1)}_{\bm{\mathrm{n}}^{+}_{bj}}(t) \\
    &\ \ \ + \sum_{ak} n_{ak} \langle \hat{f}_{ak} \hat{f}_{bj} \rangle^<_{\textsc{B}} \notag \rho^{(n-1)}_{\bm{\mathrm{n}}^{-}_{ak}}(t). \notag
\end{align}
\end{subequations}
They will be used in evaluating the commutator action of linear system-bath coupling terms. The bosonic DEOM formalism is now readily to be constructed, reading as \cite{Yan_2014, Yan_2016}
\begin{align} \label{DEOM} 
    \dot{\rho}^{(n)}_{\bm{\mathrm{n}}}(t) &= - \left( i \mathcal{L}_{\mathrm{S}} + \sum_{ak} n_{ak} \gamma_{ak} \right) \rho^{(n)}_{\bm{\mathrm{n}}}(t) - i \sum_{ak} \mathcal{Q}^{\times}_a \rho^{(n + 1)}_{\bm{\mathrm{n}}^+_{ak}}(t) \notag\\ 
    & \ \ \ - i \sum_{abk} n_{ak} \left( \eta'_{abk} \mathcal{Q}^{\times}_b + i \eta^{''}_{abk} \mathcal{Q}^{\circ}_b \right)\rho^{(n-1)}_{\bm{\mathrm{n}}^{-}_{ak}} (t) .
\end{align}
The involved superoperators and coefficients are defined as below:
\begin{align*}
    & \mathcal{L}_{\mathrm{S}} \hat{O} \equiv [\hat{H}_{\mathrm{S}}, \hat{O}], \ \ \ \ \
    \mathcal{Q}^{\times}_a \hat{O}  \equiv [\hat{Q}_{a}, \hat{O}], \ \ \ \ \ 
    \mathcal{Q}^{\circ}_a \hat{O} \equiv \{ \hat{Q}_{a}, \hat{O} \}, \\
    &\ \ \ \ \ \ \ \ \ \eta'_{abk} \equiv \frac{\eta_{abk} + \eta^*_{ab\overline{k}}}{2}, \ \ \ \ \ \eta^{''}_{abk} \equiv \frac{\eta_{abk} - \eta^*_{ab\overline{k}}}{2i}.
\end{align*}
The RK-4/RK-45 algorithm \cite{Numerical_Analysis} are usually adopted as the numerical propagation scheme of Eq.(\ref{DEOM}). There are also on-the-fly filtering algorithms available for acceleration. \cite{Filtering_2009}

\subsection{Time-domain Prony fitting decomposition}

In this context, the central problem in DEOM is to decomposite the bare-bath TCF into a sum of exponential series. Based on FDT, this can be realized by expanding-over-pole strategies, such as Matsubara spectral decomposition (MSD) \cite{Weiss} and Pad$\mathrm{\acute{e}}$ spectral decomposition (PSD) \cite{Padeoriginal_2007, Pade_2010, Pade_2011}, or various least-square fitting schemes \cite{QShi_2014, JianlanWu_2017, JianlanWu_2019, NatComm_2019}. The traditional expanding-over-poles strategies are usually restricted to certain forms of bath spectral density. Here we choose the $t$-PFD strategy. The resulting numerical efficiency of HEOM/DEOM is optimized to a great extent, especially in low temperature regimes that are usually inaccessible for other methods. 

The $t$-PFD scheme is intrinsically based on least-squares fitting algorithm, in which the real and imaginary parts of TCF are fitted separately. It is easy to extract from the bosonic FDT (c.f. Eq.(\ref{bosonic_FDT})) that 
\begin{subequations} \label{TCF_re_im}
    \begin{align}
    \mathrm{Re}[C(t)] &= \frac{1}{\pi} \int_{0}^{+\infty} d\omega\ J_{\mathrm{eff}} (\omega; \beta) \coth\left( \frac{\beta \omega}{2} \right) \cos (\omega t) \notag\\
    &= \frac{1}{\pi} \int_{0}^{+\infty} d\omega\ J(\omega) \cos (\omega t),  \label{ReCt} \\
    \mathrm{Im}[C(t)] &= \frac{1}{\pi} \int_{0}^{+\infty} d\omega\ J_{\mathrm{eff}} (\omega; \beta) \sin (\omega t) \notag\\
    &= \frac{1}{\pi} \int_{0}^{+\infty} d\omega\ J(\omega) \tanh (\beta \omega / 2) \sin (\omega t), \label{ImCt}
\end{align}
\end{subequations}
where the second line in Eq.(\ref{ReCt}) and (\ref{ImCt}) are the explicit results for $S = 1/2$ case. Intriguingly, temperature dependence is only carried by the imaginary part of TCF, in line with fermion bath but the reverse of boson bath. 

Next, we target at optimizing $K = K_r + K_i$ in
\begin{align}
    \mathrm{Re}[C(t)] = \sum_{k = 1}^{K_r} \zeta_k e^{-\lambda_k t},\ \ \ \mathrm{Im}[C(t)] = \sum_{k = 1}^{K_i} \zeta'_k e^{-\lambda'_k t},
\end{align}
where $K_r$ and $K_i$ are the number of terms in real and imaginary part fitting, respectively. We accordingly denote the $t$-PFD strategy as $K_r + K_i$. See Ref. \citenum{PFD_2022} and reference therein for detailed procedures of $t$-PFD, as well as the numerical benchmarks for several commonly used spectral density functions. We also provide more illustrations and examples on the performances of TCF fitting using $t$-PFD in Appendix \ref{appendix_C}, with respect to the most challenging SSB models we tested.

\section{Computational Details, Results and discussions} \label{section_III}

In this section, we present the numerical benchmark results of the bosonic DEOM equipped with $t$-PFD for various SSB models. The total Hamiltonian is as below:
\begin{subequations}
    \begin{align}
    \hat{H}_{\mathrm{S}} &= \epsilon \hat{\sigma}_z + \Delta \hat{\sigma}_x, \\
    \hat{h}_{\textsc{B}} &=  \sum_{j = 1}^N \omega_j \hat{s}^j_z, \\
    \hat{H}_{\mathrm{SB}} &= \hat{\sigma}_{z} \otimes \sum_{j} \sqrt{2} c_{j} \hat{s}^j_x,
\end{align}
\end{subequations}
We use the Ohmic form \cite{Caldeira_1993} with exponential cutoff for the description of the continuous bath as well as its interaction with the system,
\begin{align} \label{Ohmic_spectral_density}
    J(\omega) = \frac{\pi}{2} \alpha \omega e^{ - \omega / \omega_c}.
\end{align}
In this expression, $\alpha$ is the Kondo parameter that characterizes the system-bath coupling strength, and $\omega_c$ is the bath cut-off frequency. Since Eq.(\ref{Ohmic_spectral_density}) is defined via Eq.(\ref{spectral_density_def_original}), it is proved to be in line with Eq.(\ref{fermionic_Jdef}), according to the discussions conducted in \ref{linear_response_section} and Appendix \ref{appendix_B}.

The DEOM propagation uses the fourth-order Runge-Kutta (RK-4) integrator with time step of 0.0025/$\Delta$, together with the on-the-fly filtering algorithm \cite{Filtering_2009} with given error tolerance for acceleration. 

\subsection{Zero-temperature spin relaxation dynamics and localization}

\begin{figure*}
    \centering
    \includegraphics[width=0.9 \linewidth]{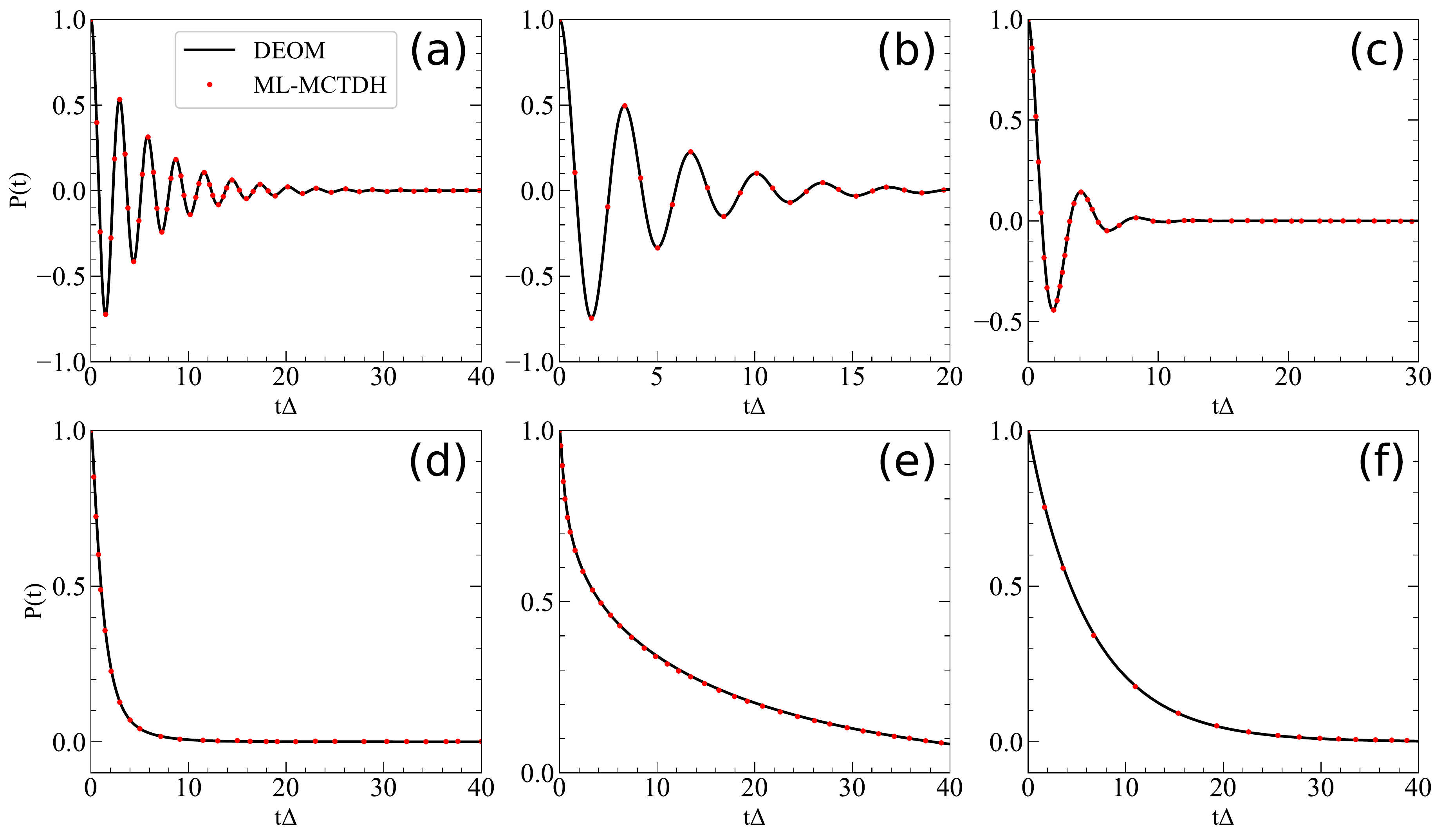}
    \caption{Population dynamics of the zero-temperature SSB models. The spin baths are parameterized as (a) $\alpha = 0.5$, $\omega_c / \Delta = 1$. (b) $\alpha = 0.1$, $\omega_c / \Delta = 6$. (c) $\alpha = 0.2$, $\omega_c / \Delta = 10$. (d) $\alpha = 0.5$, $\omega_c / \Delta = 10$. (e) $\alpha = 0.75$, $\omega_c / \Delta = 10$. (f) $\alpha = 0.5$, $\omega_c / \Delta = 40$. The ML-MCTDH results are digitized from Ref. \citenum{Wang_Shao_2012}.}
    \label{fig1}
\end{figure*}

Fig.~\ref{fig1} presents the population dynamics $P(t) \equiv \langle \hat{\sigma}_z (t) \rangle$ for zero-temperature unbiased SSB models (with $\epsilon = 0$, to keep in consistence with the previous work). Even though they equal to the spin-boson models, it provides us with the first glimpse on the power of $t$-PFD. Here we uniformly set the error tolerance of on-the-fly filtering algorithm as $5 \times 10^{-7}$. We have also confirmed that the “empirical standard” error tolerance of $1 \times 10^{-5}$ remains accurate enough in the present cases when $\alpha \leq 0.5$. For the model of Fig. \ref{fig1}e with coupling strength $\alpha = 0.75$, the convergence test suggests that the error tolerance should be set no larger than $1 \times 10^{-6}$. All the numerically converged results are obtained with $t$-PFD strategy $5+5$ (Fig. \ref{fig1}a - d, f) or $6+5$ (Fig. \ref{fig1}e) to ensure the accuracy in TCF fitting, and large enough number of tiers in the hierarchic expansion (here 20 will be satisfying). The population dynamics obtained by DEOM (black solid lines) are compared to ML-MCTDH (red dots). In all of the models presented, DEOM with $t$-PFD perfectly match the ML-MCTDH results, including the Rabi oscillations and the incoherent relaxations.

Another important phenomenon about the unbiased SSB/spin-boson model at zero temperature is the \textit{localization}, \cite{Caldeira_1993} that the population dynamics quickly reaches a biased stationary value. It is a typical phenomenon when the time scale of the bath is comparable to or longer than that of the subsystem. \cite{Phase_diagram_PRL2007, Wang_Thoss_NJP2008, Wang_Thoss_2010, JianlanWu_PRB2017} Within the adiabatic or intermediate between adiabatic and nonadiabatic regime (that with a modest $\omega_c / \Delta$ value), a large coupling strength ($\alpha > 1$) will bring about a large barrier height along the adiabatic double-well potential energy surface, such that localization of the population can be induced. \cite{Wang_Shao_2012}

\begin{figure}
    \centering
    \includegraphics[width=\linewidth]{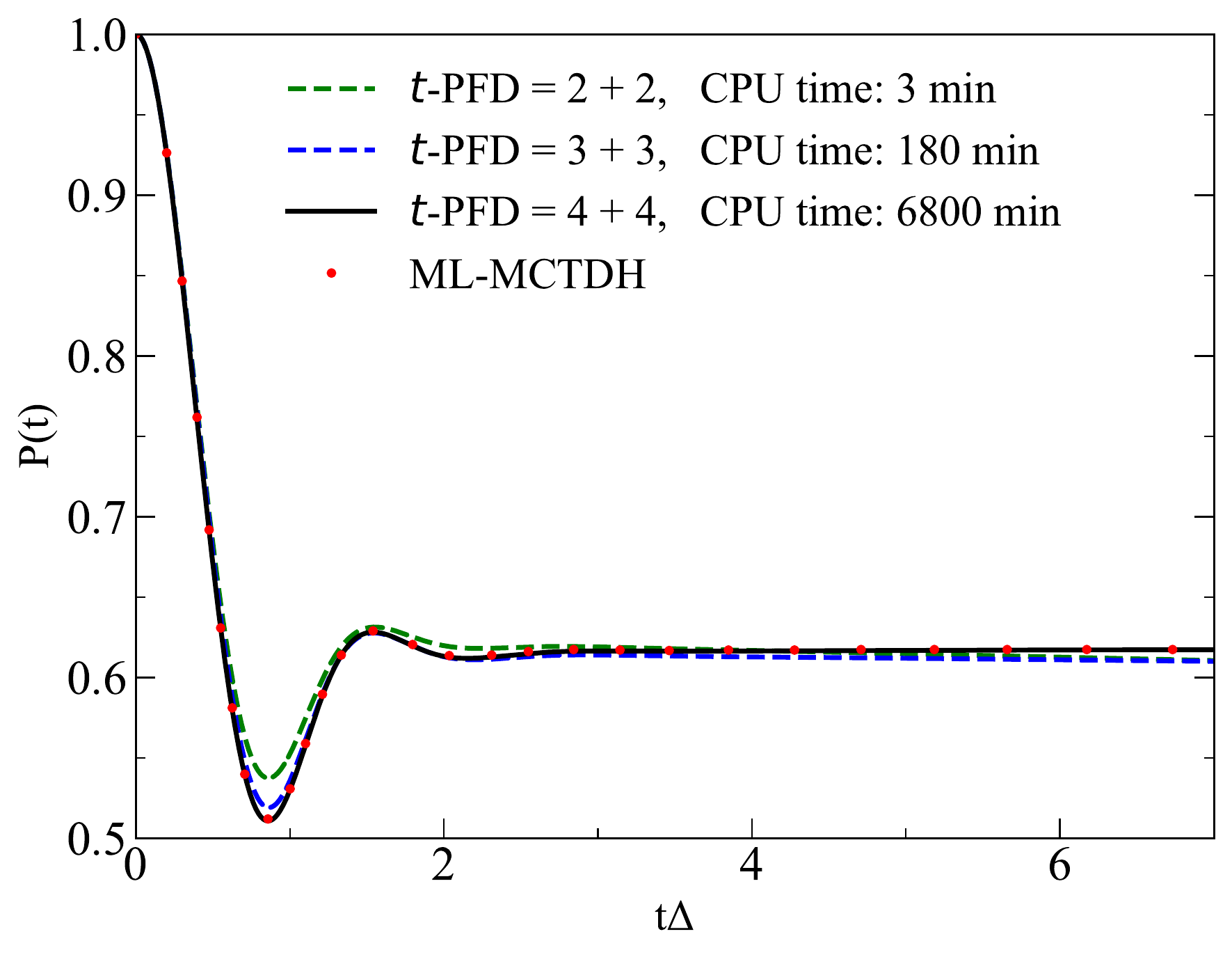}
    \caption{Convergence test of the localization model ($\omega_c / \Delta = 1,\ \alpha = 10$) using $t$-PFD strategies $2+2$, $3+3$ and $4+4$. Comparisons are made against ML-MCTDH (digitized from Ref. \citenum{Wang_Shao_2012}). }
    \label{fig2}
\end{figure}

Fig. \ref{fig2} presents the convergence test of the localization model (with $\omega_c / \Delta = 1,\ \alpha = 10$) using different $t$-PFD strategies. This computation is rather challenging, as it generally requires a lot of memory and CPU time to reach convergence. To ensure the numerical accuracy, we turn off on-the-fly filtering module, and the number of tiers is set as $45$ to stay accurate enough. As is observed that $t$-PFD strategy $2+2$ is already able to capture the localization phenomenon, but not accurate enough for population dynamics. At the expense of greater computational cost, the result generated with $t$-PFD strategy $3+3$ and $4+4$ exhibits better accuracy. It takes more than $110$ hours of CPU time (Intel Xeon Gold $6330$ CPU @$2.00$ GHz with $36$ cores) with memory requirement no less than $300$ GB to produce the $4+4$ curve. The convergence is in good agreement with ML-MCTDH. See also Appendix \ref{appendix_C} for details about the accuracy of TCF fitting under different $t$-PFD strategies. 


\subsection{Finite-temperature spin relaxation dynamics, and the localization-delocalization phase transition}

\begin{figure}
    \centering
    \includegraphics[width=\linewidth]{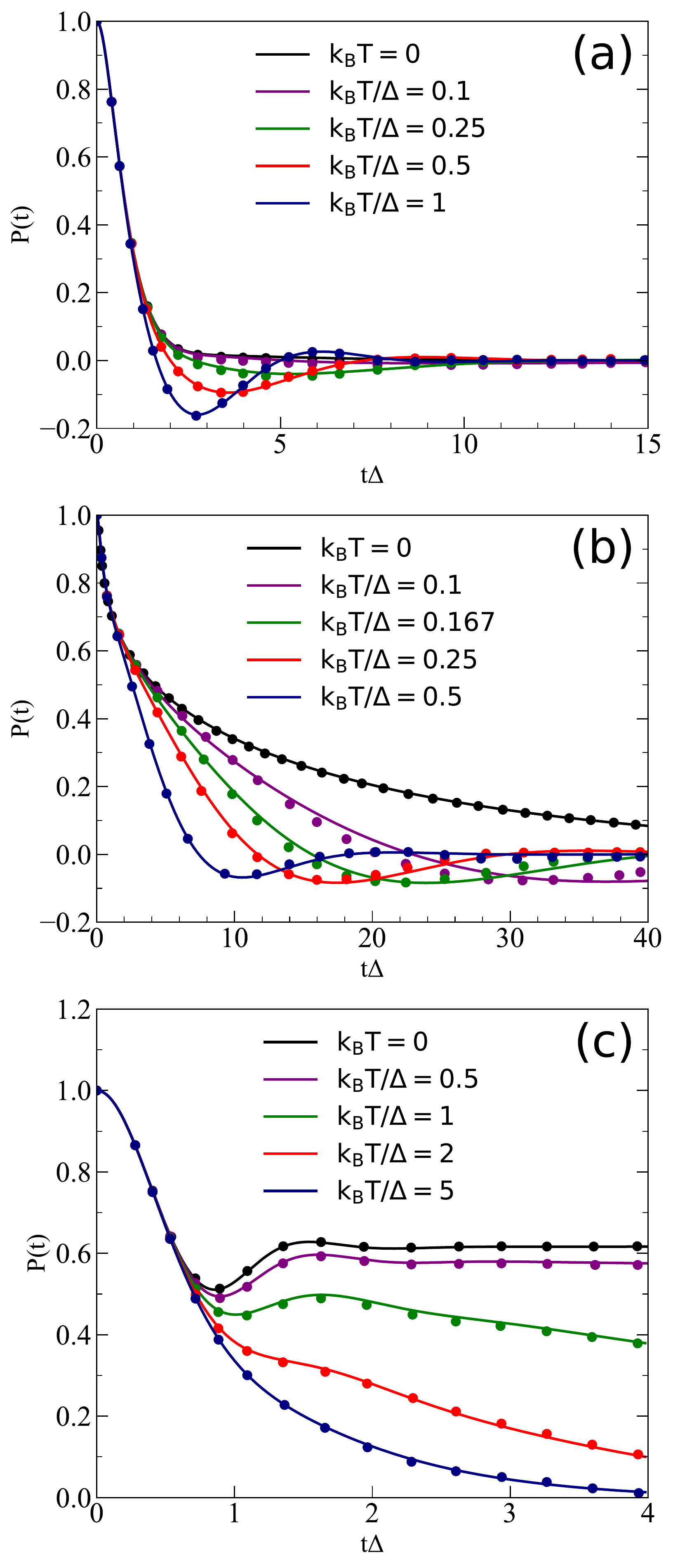}
    \caption{Temperature-dependence in the coherent-incoherent transition of population dynamics. The spin bath models are parametrized as (a) $\omega_c / \Delta = 6,\ \alpha = 0.5$. (b) $\omega_c / \Delta = 10,\ \alpha = 0.75$. (c) $\omega_c / \Delta = 1,\ \alpha = 10$. The DEOM results (solid lines) are compared to the ML-MCTDH results (solid dots, digitized from Ref. \citenum{Wang_Shao_2012}).}
    \label{fig3}
\end{figure}

Next, we turn to the finite-temperature cases. Fig. \ref{fig3} presents the temperature-dependence of spin relaxation dynamics for several representative yet challenging SSB models. 

In Fig. \ref{fig3}a, $P(t)$ under different temperatures, with bath parameters $\omega_c / \Delta = 6,\ \alpha = 0.5$ are shown (corresponding to Fig. 5a of Ref. \citenum{Wang_Shao_2012}). 
All the curves are computed after reaching a good convergence using $t$-PFD strategy $5+5$. Fig. \ref{fig3}b is similar to Fig. \ref{fig3}a but more challenging, with $\omega_c / \Delta = 10,\ \alpha = 0.75$ (corresponding to Fig. 6c of Ref. \citenum{Wang_Shao_2012}). 
So we take $t$-PFD strategy $6+5$ to reach the numerical convergence, even though $5+5$ will already be accurate enough. The comparisons are also made against ML-MCTDH. One can observe that at zero-temperature, the population dynamics exhibits clearly an incoherent decay to the equilibrium value $P = 0$. Increasing temperature by a little bit might greatly change the paradigm of decay; further increasing temperature will induce stronger coherent motion in the short time period. The temperature susceptibility is also closely related to the coupling strength $\alpha$. \cite{Wang_Shao_2012} It is well-known that at zero-temperature, the coherent-incoherent boundary is at $\alpha = 0.5$ in the scaling limit. \cite{Caldeira_1993} For the SSB model, high temperature can significantly raise up this boundary value. As is also reported by Shao and H{\"a}nggi \cite{Shao_1998}, Forsythe and Makri \cite{Makri_PRB_1999}. It is very counter-intuitive yet interesting that higher temperature will slightly enhance the coherence for the central TLS in the nonadiabatic regime ($\omega_c / \Delta \gg 1$). Physical interpretation of this abnormal phenomenon can be made from the analysis of the mapping spin-boson model with effective spectral density: $J_{\mathrm{eff}}(\omega)$ has a smaller magnitude when temperature increases, which renders weaker coupling to the central spin and win over the thermal excitation quenching effect, resulting in more coherent dynamics for the system TLS. \cite{Wang_Shao_2012} Our results are in good agreement with the previous work. 

Fig. \ref{fig3}c presents the temperature-dependence of the localization model ($\omega_c = 1,\ \alpha = 10$). 
All the curves are computed using $t$-PFD strategy $4+4$ with good convergence. As is observed that when temperature increases, the poplation distribution gets a faster decay and the bias gradually disappears, which is actually a kind of \textit{phase transition}. \cite{Caldeira_1993, Phase_transition_PRB2015} This is because increasing temperature will decrease the relaxation time of the bath, so that the system TLS become \textit{delocalized}. 

Although our results show DEOM with $t$-PFD are overall in excellent agreement with the ML-MCTDH results, we shall point out that for a certain number of finite-temperature models presented here, especially for the one with $k_B T / \Delta = 0.1$ and $0.167$ in Fig. \ref{fig3}b, there still exist minor discrepancies between DEOM and ML-MCTDH results even by eye inspection. This should be reasonable because ML-MCTDH adopts discretized bath modes as well as B-O type of approximation for high frequency bath modes, \cite{Wang_Shao_2012} it is also susceptible to the tensor network propagation scheme; while HEOM/DEOM is able to use rigorously continuous bath modes. The major resource of error for HEOM/DEOM comes from the accuracy of bath TCF fitting after reaching convergence. Another possible reason lies in the finite-temperature strategy. ML-MCTDH adopts Monte Carlo importance sampling techniques \cite{Wang_Thoss_2006} to evaluate the thermal Boltzmann operator, which is usually hard to reach numerical convergence; while HEOM/DEOM resorts to FDT, being a \textit{deterministic} pathway. 

\subsection{More results on the finite-temperature biased models}

As we know that in realistic situations, the system is not always unbiased. For example, when a Zeeman field is applied to the central spin, the energy degeneracy will be broken, which is a common experimental set up to study the radical pair spin relaxation dynamics.\cite{Manolopoulos_2019} To this reason, we further study the spin relaxation dynamics for biased models under finite-temperature, and compare them to the corresponding spin-boson model with same bath parameters.

\begin{figure*}
    \centering
    \includegraphics[width=0.75 \linewidth]{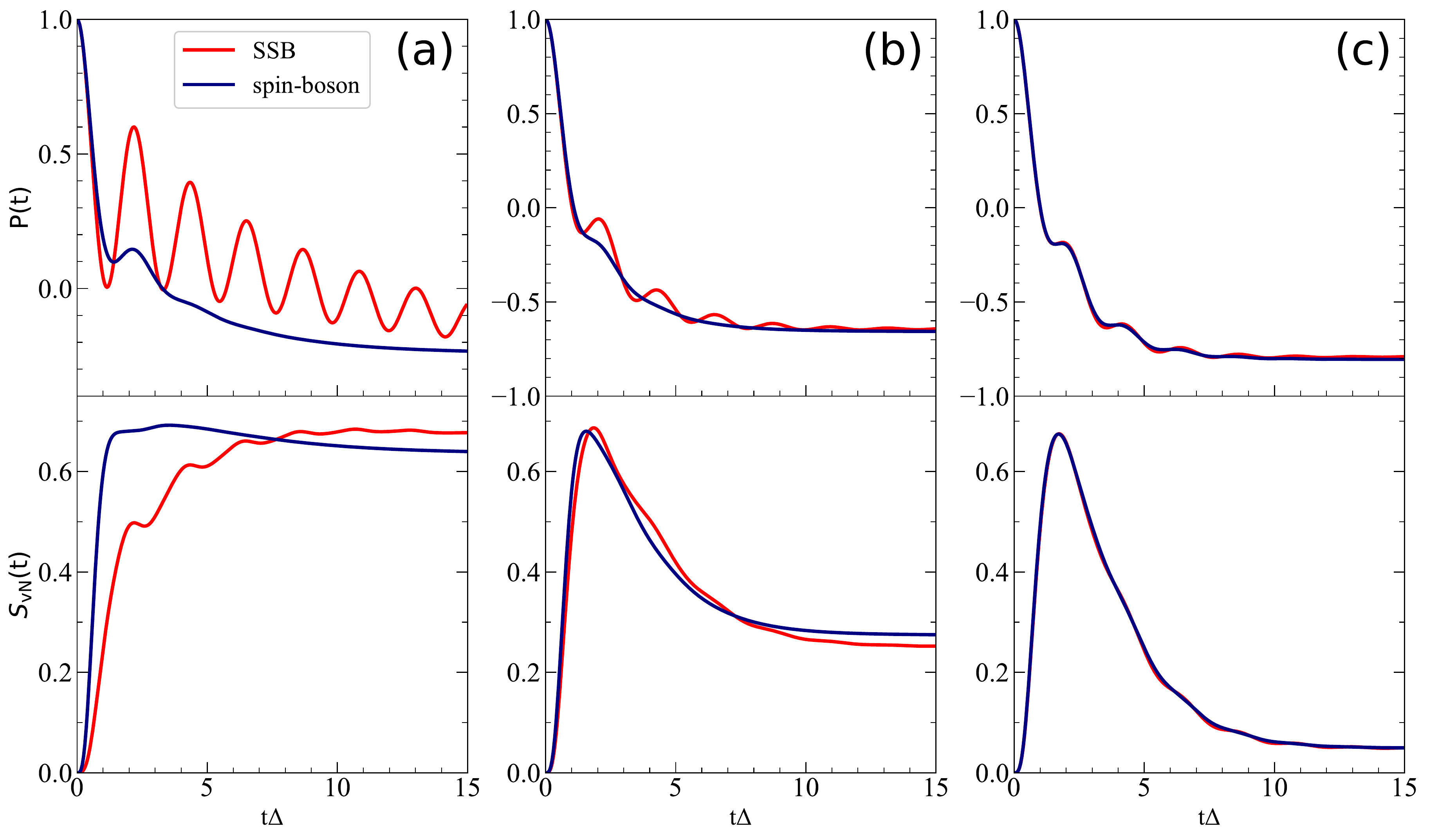}
    \caption{Population dynamics and von Neumann entropy of the biased SSB and spin-boson models. The boson/spin baths are parameterized as (a) $\alpha = 0.4$, $\omega_c / \Delta = 1$, $\beta \Delta = 0.25$. (b) $\alpha = 0.4$, $\omega_c / \Delta = 2$, $\beta \Delta = 1.0$. (c) $\alpha = 0.4$, $\omega_c / \Delta = 2$, $\beta \Delta = 5.0$. }
    \label{fig5}
\end{figure*}

Fig. \ref{fig5} shows the converged population dynamics as well as von Neumann entropy for several biased TLS (with $\epsilon = \Delta$) interacting with a spin bath (SSB) or a boson bath (spin-boson). The von Neumann entropy reads as
\begin{align}
    S_{\mathrm{vN}} (t) \equiv - \mathrm{Tr} [ \hat{\rho}_{\mathrm{S}} (t) \ln \hat{\rho}_{\mathrm{S}} (t) ], 
\end{align}
where $\hat{\rho}_{\mathrm{S}} (t)$ is the density matrix of the system TLS at time $t$. Three different models from high temperature ($\beta = 0.25$) to low temperature ($\beta = 5$) are presented here. As is observed that for high temperature model in Fig. \ref{fig5}a, the SSB model keeps much better quantum coherence than the corresponding spin-boson model. The von Neumann entropy growth is also slower. As expected, when temperature decreases, the behavior of the SSB model gradually agrees with the spin-boson model. 




\section{Concluding Remarks} \label{section_IV}

We present the numerical benchmark results of the spin relaxation dynamics of various SSB models with continuous spectral density function by using DEOM with $t$-PFD. The idea of mapping boson bath with temperature-dependent effective spectral density is validated from a microscopic perspective, and generalized to arbitrary bath spin quantum number $S$. The highlight of this strategy is that one can approximate a class of non-Gaussian bath to the Gaussian one under the linear response limit. By applying the $t$-PFD to the bath TCF, one obtains the exponential decay basis to construct DEOM and propagate. The accuracy and numerical efficiency of this strategy is illustrated by various of examples. DEOM with $t$-PFD provides an excellent agreement with the ML-MCTDH results reported in literature, despite the rather challenging model parameters. Moreover, comparing to previous research work on the SSB model based on ML-MCTDH, modular path integral, \textit{etc.}, DEOM equipped with $t$-PFD has the advantage that the bath spectral density is rigorously continuous, no B-O type of approximation is needed; and the finite-temperautre strategy is based on FDT rather than stochastic sampling, thus allowing long time propagation. In summary, it serves as a novel and more efficient benchmark scheme for quantum dynamics with a spin bath. 

The presented theoretical work provide new insights to extend HEOM/DEOM to certain types of non-Gaussian environments with arbitrary bath TCFs or spectral density functions. Future research work on methods development shall be carried out on more complicated types of interacting system-bath models that listed in Eq.(\ref{HSB_choices}), with incorporation of quadratic or higher order system-bath couplings, as well as the nonlinear response effects; we would also like to seek for applications in the simulation of radical pair spin relaxation dynamics that exposed to an enviroment of nuclear spins. 

\begin{acknowledgments}
This work was supported by the National Science Foundation CAREER Award under Grant No. CHE-1845747. Computing resources were provided by the Center for Integrated Research Computing (CIRC) at the University of Rochester. W. Y. would like to sincerely appreciate YiJing Yan for his meticulous guidance when he was at USTC, and for his valuable comments on this manuscript. 
\end{acknowledgments}

\section*{Data Availability}
The data that support the findings of this work are available from the corresponding author under reasonable request.

\appendix

\section{FDT for the spin bath, and the effective bosonic environment} \label{appendix_B}

\subsection{$S = 1/2$ case}

For spin-1/2 particles, we adopt the well-known Jordan-Wigner correspondence \cite{JordanWigner_1928} for the spin operators:
\begin{subequations}
    \begin{align} \label{JordanWigner}
    \hat{d}_j^{\dagger} &\equiv \hat{s}^j_x + i \hat{s}^j_y, \\
    \hat{d}_j &\equiv \hat{s}^j_x - i \hat{s}^j_y, \\ 
    \hat{s}^j_z &= \hat{d}_j^{\dagger}\hat{d}_j - \frac{1}{2}.
\end{align}
\end{subequations}
where $\hat{d}_j$, $\hat{d}_j^{\dagger}$ are the mapping \textit{fermionic} annihilation/creation operators with anti-commutation relation on the same site. However, on different sites we have bosonic commutation relations, which means spins on different sites commute, unlike fermions which anti-commute. For this reason, spins are also referred to as \textit{hard-core bosons} or \textit{spinless fermions}, \cite{Girardeau_1960} whose commutation relations can be summarized as
\begin{subequations} \label{hardcoreboson}
    \begin{align}
    & \{ \hat{d}_j, \hat{d}_j \} = \{ \hat{d}_j^{\dagger}, \hat{d}_j^{\dagger} \} = 0,\ \ \ \{ \hat{d}_j, \hat{d}_j^{\dagger} \} = 1; \\
    & \{ \hat{d}_j, \hat{d}_k \} = 2 \hat{d}_j \hat{d}_k,\ \ \ \{ \hat{d}^{\dagger}_j, \hat{d}^{\dagger}_k \} = 2 \hat{d}^{\dagger}_j \hat{d}^{\dagger}_k,\notag\\
    & \{ \hat{d}_j, \hat{d}^{\dagger}_k \} = 2 \hat{d}_j \hat{d}^{\dagger}_k,\ \ \  k\neq j; \\
    & [\hat{d}_j, \hat{d}_j ] = [ \hat{d}_j^{\dagger}, \hat{d}_j^{\dagger} ] = 0,\ \ \ [\hat{d}_j, \hat{d}_j^{\dagger}] = 1 - 2\hat{d}_j^{\dagger}\hat{d}_j; \\
    & [\hat{d}_{j}, \hat{d}_{k}] = [\hat{d}_{j}^{\dagger}, \hat{d}_{k}^{\dagger}] = [\hat{d}_{j}, \hat{d}_{k}^{\dagger}] = 0,\ \ \ k\neq j.
\end{align}
\end{subequations}

As is shown that the hard-core bosons possess characters of both bosons and fermions, one can either define the linear response functions using commutator or the single particle Green's function using anti-commutator. Here we will show both possibilities. For convenience, let's first write down the time evolution of single bath operators, 
\begin{align} \label{d_ddagger_t}
    \hat{d}_j^{\dagger}(t) &= e^{i\hat{h}_{\textsc{B}}t} \hat{d}_j^{\dagger}(0) e^{ - i\hat{h}_{\textsc{B}}t} = \hat{d}_j^{\dagger} (0) e^{ i \omega_j t}, \notag\\
    \hat{d}_j(t) &= e^{i\hat{h}_{\textsc{B}}t} \hat{d}_j(0) e^{ - i\hat{h}_{\textsc{B}}t} = \hat{d}_j (0) e^{- i \omega_j t},
\end{align}
where we have used the well-known Baker-Campbell-Hausdorff identity. \cite{GTM222}

We will start with the anti-commutator version. For each pair of mapping spinless fermionic creation/annihilation operators $\hat{d}_j^{\dagger}$ and $\hat{d}_j$ that satisfy the commutation relations defined in Eq.(\ref{hardcoreboson}), the single particle Green's functions can be defined and evaluated as
\begin{align}
    \langle \{ \hat{d}_j(t), \hat{d}^{\dagger}_k(0) \} \rangle_{\textsc{B}} &= \delta_{jk} e^{- i \omega_j t}, \notag\\ 
    \langle \{ \hat{d}^{\dagger}_j(t), \hat{d}_k(0) \} \rangle_{\textsc{B}} &= \delta_{jk} e^{i \omega_j t}, \notag\\
    \langle \{ \hat{d}_j(t), \hat{d}_k(0) \} \rangle_{\textsc{B}} &= \langle \{ \hat{d}^{\dagger}_j(t), \hat{d}^{\dagger}_k(0) \} \rangle_{\textsc{B}} = 0.
\end{align}
So we have
\begin{align} \label{fermionic_lrf}
    \langle \{ \hat{F}(t), \hat{F}(0) \} \rangle_{\textsc{B}} = \sum_j c^2_j \cos(\omega_j t).
\end{align}
As a result, the spectral density function is evaluated as (c.f. Eq.(\ref{fermionic_Jdef}))
\begin{align} \label{spectraldensity_fermionic_lsr}
    J'(\omega) &= \frac{1}{2} \int_{- \infty}^{+\infty} dt\ e^{i \omega t} \langle \{\hat{F}(t), \hat{F}(0) \} \rangle_{\textsc{B}} \notag\\
    &= \frac{\pi}{2} \sum_j c^2_j [\delta(\omega - \omega_j) + \delta(\omega + \omega_j)],
\end{align}
which is an extension of the result given by Caldeira and Leggett \cite{Caldeira_Leggett_1983} (see also Eq.(\ref{spectral_density_def_original})) to negative frequencies $\omega < 0$, while ensuring it to be an even function.

The derivation via the commutator version is very similar to the anti-commutator one. Following the same procedure, one can easily obtain the bare-bath linear response function as
\begin{align}
    i \langle [\hat{F}(t), \hat{F}(0)] \rangle_{\textsc{B}} = \sum_j c^2_j \langle (1 - 2\hat{d}_j^{\dagger}\hat{d}_j) \rangle_{\textsc{B}} \sin(\omega_j t).
\end{align}
For independent spin $S = 1/2$ particles, we have
\begin{align} \label{partition_function}
    Z_{\textsc{B}} \equiv \mathrm{Tr}_{\textsc{B}} [ e^{- \beta \hat{h}_{\textsc{B}}} ] = \prod_j 2 \cosh \left( \frac{\beta \omega_j}{2} \right),
\end{align}
one immediately obtains
\begin{align} \label{bosonic_lrf}
    i \langle [\hat{F}(t), \hat{F}(0)] \rangle_{\textsc{B}} = \sum_j c^2_j \tanh \left( \frac{\beta \omega_j}{2} \right) \sin(\omega_j t).
\end{align}
And the corresponding spectral density function reads as (c.f. Eq.(\ref{spectral_density_def}))
\begin{align} \label{Jeff2}
    J(\omega) &\equiv \frac{1}{2} \int_{- \infty}^{+ \infty} dt\ e^{i \omega t} \langle [\hat{F}(t), \hat{F}(0)] \rangle_{\textsc{B}} \\
    &= \frac{\pi}{2} \sum_j c^2_j \tanh \left( \frac{\beta \omega_j}{2} \right) [\delta(\omega - \omega_j) - \delta(\omega + \omega_j)], \notag
\end{align}
which is an odd function. On the other hand, we can rewrite Eq.(\ref{Jeff2}) as
\begin{align} \label{Jeff2_re}
    J(\omega) &= \frac{\pi}{2} \sum_j c^2_j \tanh \left( \frac{\beta \omega}{2} \right) [\delta(\omega - \omega_j) + \delta(\omega + \omega_j)] \notag\\
    &= J'(\omega) \tanh \left( \frac{\beta \omega}{2} \right),
\end{align}
where $J'(\omega)$ is defined in Eq.(\ref{spectraldensity_fermionic_lsr}), giving rise to the temperature-dependent effective spectral density.

Combining Eq.(\ref{fermionic_lrf}) and (\ref{bosonic_lrf}), one obtains
\begin{align} \label{FDT_proof}
    C(t) &\equiv \langle \hat{F}(t) \hat{F}(0) \rangle_{\textsc{B}} = \frac{1}{2} \sum_j c^2_j \left[ \frac{e^{ - i \omega_j t}}{1 + e^{- \beta \omega_j}} + \frac{ e^{i \omega_j t} }{1 + e^{ \beta \omega_j}} \right] \notag\\
    &= \frac{1}{2} \sum_j c^2_j [\delta(\omega - \omega_j) + \delta(\omega + \omega_j) ] \frac{e^{ - i \omega t}}{1 + e^{- \beta \omega}} \notag\\
    &= \frac{1}{\pi} \int_{- \infty}^{+\infty} d \omega \frac{e^{- i \omega t} J'(\omega) }{1 + e^{- \beta \omega}},
\end{align}
which is just Eq.(\ref{fermionic_FDT}) in the main text, the FDT for spin bath. It has an equivalent bosonic FDT formalism if one takes $J_{\mathrm{eff}} (\omega; \beta) \equiv J(\omega) = J'(\omega) \tanh(\beta\omega / 2)$, giving rise to Eq. (\ref{bosonic_FDT}). Thus, the spin-boson problem with an effective spectral density function arises naturally.

An alternative but similar argument can be done by using the coupled-fermion representation for spin operators ($S = 1/2$), discussed by Mattis, et al.,\cite{Mattis, Callen_1966, Cao_2018_2} reading as
\begin{subequations}
    \begin{align} \label{coupled_fermion}
    \hat{s}^j_+ &= \hat{c}_j^{\dagger}(\hat{d}_j + \hat{d}_j^{\dagger}), \\
    \hat{s}^j_- &= (\hat{d}_j + \hat{d}_j^{\dagger}) \hat{c}_j, \\ 
    \hat{s}^j_z &= \hat{c}_j^{\dagger}\hat{c}_j - \frac{1}{2},
\end{align}
\end{subequations}
where $\hat{c}, \hat{c}^{\dagger}$ and $\hat{d}, \hat{d}^{\dagger}$ are two sets of fermion operators that anti-commute with each other. On the other hand, since the mapping fermionic bath operators still commute with the system operators rather than anti-commute, it should also lead to the same bosonic DEOM formalism, as is studied by Jin, et al. \cite{JSJ_2007}  based on the influence functional. 

\begin{figure}
    \centering
    \includegraphics[width=0.9\linewidth]{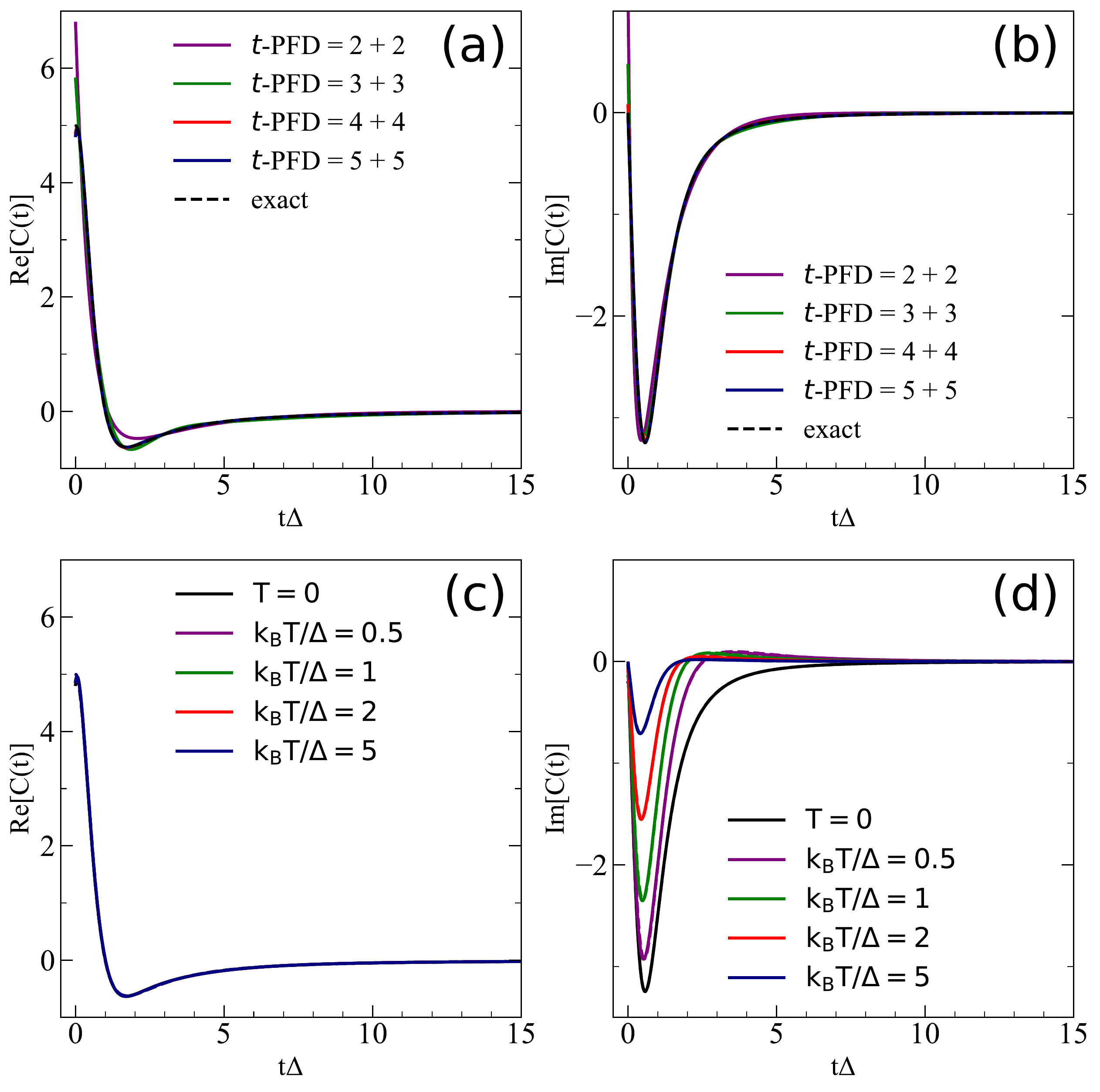}
    \caption{$t$-PFD fitting results for the spin bath with $\omega_c / \Delta = 1,\ \alpha = 10$. (a), (b) real and imaginary part fitting results for the zero-temperature model using 2, 3, 4 and 5 terms, respectively. (c), (d) real and imaginary part fitting results for different finite-temperature models using 4 terms; the dashed lines represent the exact TCF, and the solid lines represent the fitting results, which are almost overlapped. }
    \label{fig4}
\end{figure}

\subsection{Generalization to arbitrary spin $S$}
Adopting the same angular momentum raising/lowering operators that are defined as
\begin{align}
    \hat{s}^j_+ \equiv \hat{s}^j_x + i \hat{s}^j_y,\ \ \ \ \ 
    \hat{s}^j_- \equiv \hat{s}^j_x - i \hat{s}^j_y,
\end{align}
so that the original $\otimes_{j = 1}^N \mathfrak{su}(2)$ Lie algebra becomes
\begin{align}
    [\hat{s}^i_+, \hat{s}^j_-] = 2 \hat{s}^i_z \delta_{ij}, \ \ \ \ \ [\hat{s}^i_z, \hat{s}^j_{\pm}] = \pm \hat{s}^i_{\pm} \delta_{ij}.
\end{align}
Their time dependence can be evaluated as
\begin{align}
    \hat{s}^j_{\pm} (t) = e^{i\hat{h}_{\textsc{B}}t} \hat{s}^j_{\pm} (0) e^{ - i\hat{h}_{\textsc{B}}t} = \hat{s}^j_{\pm} (0) e^{\pm i \omega_j t}. 
\end{align}
The bare-bath partition function can be evaluated as
\begin{align}
    Z_{\textsc{B}} = \prod_j \left(2 \sum_{k = 0}^{[S]} \cosh\left( (k + \{S\}) \beta \omega_j \right) - \Delta(S) \right),
\end{align}
where $[S]$, $\{ S \}$ are the integer and fractional part of $S$, with $S = [S] + \{S\}$; $\Delta(S) = 1$ for $S$ being integers, and $\Delta(S) = 0$ for $S$ being half integers. As a result,
\begin{align} \label{Sz_expression}
    \langle \hat{s}^j_z \rangle_{\textsc{B}} = - \frac{\partial \ln Z_{\textsc{B}}}{\partial (\beta \omega_j)}, \ \ \ \ \ 
    \langle (\hat{s}^j_z)^2 \rangle_{\textsc{B}} = \frac{\partial^2 \ln Z_{\textsc{B}}}{\partial (\beta \omega_j)^2 }.
\end{align}
It is also easy to obtain the bare-bath single particle Green's function and linear response function as:
\begin{subequations}
    \begin{align} 
    \langle \{ \hat{F}(t), \hat{F}(0) \} \rangle_{\textsc{B}} &= \frac{1}{S} \sum_j c^2_j [ S(S+1) - \langle (\hat{s}^j_z)^2 \rangle_{\textsc{B}} ] \cos (\omega_j t), \\
    i \langle [ \hat{F}(t), \hat{F}(0) ] \rangle_{\textsc{B}} &= - \frac{1}{S} \sum_j c^2_j \langle \hat{s}^j_z \rangle_{\textsc{B}} \sin (\omega_j t). \label{bosonic_lrf_2}
\end{align}
\end{subequations}
And the corresponding spectral density functions can be evaluated by Eq.(\ref{spectral_density_def}) and (\ref{fermionic_Jdef}) as
\begin{subequations}
    \begin{align} 
    J'(\omega) &= \frac{\pi}{2S} \sum_j c^2_j [ S(S+1) - \langle (\hat{s}^j_z)^2 \rangle_{\textsc{B}} ] \notag\\
    &\ \ \ \ \ \times [\delta(\omega - \omega_j) + \delta(\omega + \omega_j)], \label{spectraldensity_fermionic_lsr_2} \\
    J(\omega) &= - \frac{\pi}{2S} \sum_j c^2_j \langle \hat{s}^j_z \rangle_{\textsc{B}} [\delta(\omega - \omega_j) - \delta(\omega + \omega_j)]. \label{spectraldensity_bosonic_lsr_2}
\end{align}
\end{subequations}
They are derived from a microscopic perspective, satisfying the correct symmetry; however, neither of them give rise to the original definition of bath spectral density function that given by Caldeira and Leggett \cite{Caldeira_Leggett_1983}. One might choose to still using Eq.(\ref{spectral_density_def_original}) and directly getting $J(\omega)$ in Eq.(\ref{spectraldensity_bosonic_lsr_2}) as the effective spectral density, or do modifications to make it in accordance with Eq.(\ref{spectraldensity_fermionic_lsr_2}) or (\ref{spectraldensity_bosonic_lsr_2}). For example, if we use (c.f. Eq.(\ref{spectraldensity_fermionic_lsr_2})):
\begin{align}
    J'(\omega) &\equiv \frac{\pi}{2S} \sum_j c^2_j [ S(S+1) - \langle (\hat{s}^j_z)^2 \rangle_{\textsc{B}} ] \Big|_{\omega_j = \omega} \notag\\
    &\ \ \ \ \ \times [\delta(\omega - \omega_j) + \delta(\omega + \omega_j)],
\end{align}
where the footnote $\omega_j = \omega$ means replacing all the $\omega_j$ in the prefactors by $\omega$. Then the theory of effective spectral density function can arise as
\begin{align}
    C(t) &\equiv \langle \hat{F}(t) \hat{F}(0) \rangle_{\textsc{B}} = \frac{1}{\pi} \int_{- \infty}^{+\infty} d \omega e^{- i \omega t} \frac{J_{\mathrm{eff}}(\omega; \beta, S)}{1 - e^{- \beta \omega}}, \notag\\
    & J_{\mathrm{eff}}(\omega; \beta, S) \equiv J(\omega) = J'(\omega) \zeta(\omega; \beta, S),
\end{align}
with 
\begin{align}
    \zeta(\omega; \beta, S) &\equiv \frac{1 - e^{- \beta \omega}}{2} \\
    &\ \ \ \times \frac{S(S+1) - \langle (\hat{s}^j_z)^2 \rangle_{\textsc{B}} - \langle \hat{s}^j_z \rangle_{\textsc{B}}}{S(S+1) - \langle (\hat{s}^j_z)^2 \rangle_{\textsc{B}}} \Bigg|_{\omega_j = \omega}. \notag
\end{align}
One can check that $\zeta(\omega; \beta, S = 1/2) = \tanh(\beta\omega / 2)$. 

On the other hand, under the high spin limit of $S \gg 1$, Eq.(\ref{spectraldensity_bosonic_lsr_2}) will reduce to the conventional form, 
\begin{align} \label{highspin_J}
    J(\omega) = \frac{\pi}{2} \sum_j c^2_j [\delta(\omega - \omega_j) - \delta(\omega + \omega_j)],
\end{align}
which is in line with the original definition in Eq.(\ref{spectral_density_def_original}) but with odd analytical continuation. As a result, the spin bath under the high spin limit is isomorphic to the boson bath. This can be understood via the Holstein-Primakoff transformation, \cite{Shankar_2012} 
\begin{align}
    \hat{s}^{i}_{+} &= \sqrt{2S}\ \sqrt{1 - \frac{\hat{b}^{\dagger}_i \hat{b}_i}{2S}}\ \hat{b}_i \approx \sqrt{2S}\ \hat{b}_i,  \notag\\
    \hat{s}^{i}_{-} &= \sqrt{2S}\ \hat{b}^{\dagger}_i\ \sqrt{1 - \frac{\hat{b}^{\dagger}_i \hat{b}_i}{2S}} \approx \sqrt{2S}\ \hat{b}^{\dagger}_i, \ \notag\\
    \hat{s}^{i}_{z} &= S - \hat{b}^{\dagger}_i \hat{b}_i,
\end{align}
where $\hat{b}^{\dagger}_i$ and $\hat{b}_i$ are bosonic creation/annihilation operators that satisfy the Heisenberg commutation relations. Consequently,
\begin{align} \label{highspin_F}
    \hat{F} = \frac{1}{\sqrt{S}} \sum_j c_{j} \hat{s}^j_x \approx \sum_j \frac{c_{j}}{\sqrt{2}} (\hat{b}_j + \hat{b}^{\dagger}_j) \equiv \sum_j c_{j} \hat{x}_j,
\end{align}
where $\hat{x}_j \equiv (\hat{b}_j + \hat{b}^{\dagger}_j) / \sqrt{2}$. Eq.(\ref{highspin_F}) recovers the bath dissipation operator of the conventional spin-boson model. 

One should find it straightforward to generalize the discussions above to more complicated interacting system-bath models, which might contain multiple dissipation modes as is listed in Eq.(\ref{HSB_choices}), by carrying out the very similar arguments. The spectral density functions will be anistropic in such cases. Even more general, the linear response limit can be easily applied to general finite baths \cite{Finite_Bath, Suarez_Silbey_1991}, with multiple dissipation modes. \textit{i.e.}, the individual bath particles are general level systems with $SU(N)$ symmetry. In these situations, the generators of $\mathfrak{su}(N)$ Lie algebra \cite{Richardson_2020, Duncan_2022} can be applied. Another pathway to establish the theory of effective spectral density function could be achieved by using the generalized Schwinger's theory of angular momentum, \footnote{J. Schwinger, in \textit{Quantum Theory of Angular Momentum}, edited by L. C. Biedenharn and H. Van Dam (Academic, New York, 1965); see also J. J. Sakurai, \textit{Modern Quantum Mechanics} (Addison-Wesley, New York, 1994), p. 217.} which remains to be further explored.

\section{The numerical accuracy of $t$-PFD strategies} \label{appendix_C}

In this section, we will provide examples of different $t$-PFD strategies to illustrate its numerical accuracy on TCF fitting. 

Fig. \ref{fig4} presents the $t$-PFD results for the real and imaginary parts of bare-bath TCFs. We take the most challenging model with bath parameters $\omega_c / \Delta = 1,\ \alpha = 10$, and the bare-bath TCF \textit{plateau time} is taken as $40 \Delta$ with resolution $dt = 0.01 \Delta$ uniformly. \cite{PFD_2022} Fig. \ref{fig4}a and b present the accuracy of fitting for the zero-temperature model using different number of terms. As is seen that using 4 or 5 terms will be accurate enough to fit the real and imaginary parts of the TCF, respectively. Fig. \ref{fig4}c and d present the fitting results of finite-temperature models using $4+4$ 
scheme. As is observed straightforwardly that the real part of TCF is temperature-independent, only the imaginary part varies with temperature, in accordance with Eq. (\ref{TCF_re_im}). The finite-temperature models are expected generally more difficult to be accurately fitted than the zero-temperature model. Here all curves are accurately fitted using $4+4$ scheme by eye inspection.

In summary, to reach better accuracy, one will need to use more terms, but the expense is that the computational cost grow drastically. For practical use, one will need to explore the proper $t$-PFD strategy to balance accuracy and computational cost with regarding to the specific bath types and parameters.

\providecommand{\noopsort}[1]{}\providecommand{\singleletter}[1]{#1}%

\end{document}